\pdfoutput=1
\documentclass[conference]{IEEEtran}
\IEEEoverridecommandlockouts
\usepackage{cite}
\usepackage{amsmath,amssymb,amsfonts}
\usepackage{algorithmic}
\usepackage{graphicx}
\usepackage{textcomp}
\usepackage{xcolor}
\def\BibTeX{{\rm B\kern-.05em{\sc i\kern-.025em b}\kern-.08em
    T\kern-.1667em\lower.7ex\hbox{E}\kern-.125emX}}
\usepackage{tabularx}
\usepackage{multirow}
\usepackage{footnote}
\usepackage{placeins}
\usepackage{hyperref}
\usepackage{amssymb}
\usepackage{amsmath}
\usepackage{amsfonts}
\hypersetup{
  colorlinks = true,
  linkcolor  = blue,
  urlcolor   = blue,
  citecolor  = black,
}
\usepackage{url}
\usepackage{cleveref}
\usepackage{thmbox}
\usepackage{mdframed}
\usepackage{dblfloatfix}

\newcommand*{\fullref}[1]{\hyperref[{#1}]{\cref*{#1} \nameref*{#1}}}
\newcommand*{\Fullref}[1]{\hyperref[{#1}]{\Cref*{#1} \nameref*{#1}}}
\newcommand*{\secref}[1]{\hyperref[{#1}]{\autoref*{#1}}}
\newcommand*{\Secref}[1]{\hyperref[{#1}]{\Cref*{#1}}}

\let\oldFootnote\footnote
\newcommand\nextToken\relax

\renewcommand\footnote[1]{%
    \oldFootnote{#1}\futurelet\nextToken\isFootnote}

\newcommand\isFootnote{%
    \ifx\footnote\nextToken\textsuperscript{,}\fi}

\usepackage{booktabs}
\usepackage{xcolor}
\def\BibTeX{{\rm B\kern-.05em{\sc i\kern-.025em b}\kern-.08em
    T\kern-.1667em\lower.7ex\hbox{E}\kern-.125emX}}

\newcounter{openboxwithtitle}[section]
\newenvironment{openboxwithtitle}[1]{
  \refstepcounter{openboxwithtitle} 
  \vspace{1.75ex}
  \thmbox[L]{\textbf{#1}}
  \hspace*{-1.5em}\slshape\ignorespaces%
}{
  \endthmbox\vspace*{.75ex}
}

\begin{document}

\title{SciCom Wiki: A Digital Library to Support the Science Communication Knowledge Infrastructure for Videos and Podcasts 
\thanks{This work was co-funded by the DFG SE2A Excellence Cluster, as well as the NFDI4Ing project funded by the German Research Foundation (project number 442146713) and NFDI4DataScience (project number 460234259).}
}

\author{
    \IEEEauthorblockN{
    Tim Wittenborg\IEEEauthorrefmark{1},
    Niklas Stehr,
    Oliver Karras\IEEEauthorrefmark{2},
    Sören Auer\IEEEauthorrefmark{1}\IEEEauthorrefmark{2}}
    \IEEEauthorblockA{\IEEEauthorrefmark{1}L3S Research Center, Leibniz University Hanover, Hanover, Germany
    \\tim.wittenborg@l3s.uni-hannover.de}
    \IEEEauthorblockA{\IEEEauthorrefmark{2}TIB - Leibniz Information Centre for Science and Technology, Hanover, Germany
    \\oliver.karras@tib.eu, soeren.auer@tib.eu}
}

\maketitle

\begin{abstract}
Videos and Podcasts have established themselves as the medium of choice for civic dissemination, but also as carriers of misinformation.
The emerging Science Communication Knowledge Infrastructure (SciCom KI), which curates these increasingly non-textual media, remains fragmented and inadequately equipped to scale against the content flood.
Our work sets out to support the SciCom KI with a central, collaborative platform, the SciCom Wiki,
to facilitate FAIR (findable, accessible, interoperable, reusable) media representation, particularly for videos and podcasts.
We survey requirements from 53 stakeholders and individually refine these insights in 11 interviews.
We then design and implement an open-source service system centered on Wikibase and evaluate our prototype with another 14 participants.
Overall, our findings identified several needs to support the SciCom KI systematically.
Our SciCom Wiki approach was found suitable to address the raised requirements.
Further, we identified that the SciCom KI is severely underdeveloped regarding FAIR knowledge and related systems facilitating its collaborative creation and curation.
Our system can provide a central knowledge node similar to Wikidata, yet a collaborative effort is required to scale the necessary features against the imminent (mis-)information flood.
\end{abstract}

\begin{IEEEkeywords}
knowledge infrastructure, science communication, digital library, audiovisual content, wiki, media annotation
\end{IEEEkeywords}

\section{Introduction}
In today's digital landscape, freely accessible scientific videos and podcasts exist alongside unreliable media~\cite{hendriks_measuring_2015}.
Users face challenges identifying valuable knowledge amidst this misinformation.
While platforms from YouTube to Twitch to TikTok  are committed to combat disinformation\footnote{\url{https://digital-strategy.ec.europa.eu/en/library/signatories-2022-strengthened-code-practice-disinformation}}, they still provide limited tools for effective fact-checking, contextualization or criticism~\cite{hussein_measuring_2020}.
There are various endeavors to address these issues:
dedicated channels sprout in the same environments to fact-check their peers~\cite{lu_unpacking_2023}, such as Quarks Science Cops\footnote{\url{https://www.youtube.com/@quarkssciencecops}} or Doktor Whatson\footnote{\url{https://www.youtube.com/@DoktorWhatson}};
organizations form to highlight specific videos or podcasts, such as the world lecture project\footnote{\url{https://world-lecture-project.org/}} or Wissen\{schaft\}spodcasts\footnote{\url{https://wissenschaftspodcasts.de/}};
scientists analyze scientific podcasts specifically, such as McKenzie~\cite{mackenzie_science_2019} or Wimpact\cite{kuhle_wi4impact_2025},
or even dedicated platforms rising to provide reviewed alternatives to the commercialized monoliths, such as the AV-Portal~\cite{marin_arraiza_tibav_2015}.
What all these curation projects lack is a central infrastructure to unify their knowledge.
While Knowledge Infrastructures (KIs) like Wikipedia, Wikidata, and Wikisource successfully curate textual information at scale~\cite{giles_internet_2005}, even their resources and scalability are limited\footnote{\url{https://www.wikidata.org/wiki/Wikidata:WikiProject_Limits_of_Wikidata}}.
Their broad scope and high notability criteria exclude a wide range of scientifically relevant media, resulting in structural gaps.
Wikimedia has identified these gaps and started the Wikibase Ecosystem\footnote{\url{https://wikiba.se/}}~\cite{varvantakis_wikibase_2025}, sprouting hundreds of dedicated knowledge bases, like MiMoText~\cite{schoch_smart_2022} for literary history and literary historiography.
Despite this, modern non-textual media (NTM) such as video and audio remain largely unaddressed.

Our work seeks to answer these needs with an open digital library for scientifically relevant audiovisual media.
We present a wiki-based system providing an index of information on scientific videos and podcasts, enabling users to add, edit, and curate content as established in the Wikiverse. 
In addition, the system promotes participation in further development, providing open interfaces to utilize this collaborative database in various future use cases.
This approach aims to provide a uniform digital infrastructure built upon the principles of open source and open access as well as \textbf{F}indable, \textbf{A}ccessible, \textbf{I}nteroperable and \textbf{R}eusable (FAIR) data~\cite{wilkinson_fair_2016}.
Our contributions include:

\textbf{1.)} An overview of the current state of Science Communication Knowledge Infrastructure (SciCom KI), including requirements gathered via 53 anonymous survey participants and 11 stakeholder interviews.

\textbf{2.)} The design and implementation of a digital library and interface to provide tool- and infrastructure-support for collaborative media representation, processing, and annotation.

\textbf{3.)} An evaluation by 14 stakeholders, confirming that the system has a good user experience and is capable of meeting several requirements already, while extensive future work is required to address them all eventually.

This work is structured as follows: 
\Secref{sec:related} presents the background and related work on SciCom, KI, and SciCom KI.
\Secref{sec:approach} illustrates the approach, including the requirements' elicitation, which informed our implementation of the digital library described in \secref{sec:implementation}.
The evaluation of the current implementation is detailed in \secref{sec:evaluation}, including the limitations.
These findings are discussed in \secref{sec:discussion}, highlighting future work and concluding in \secref{sec:conclusion}.

\section{Background and Related Work\label{sec:related}}
We summarize the fundamentals of \nameref{sec:ki} and \nameref{sec:scicom} to focus on their convergence in \nameref{sec:scicomki}.

\subsection{Knowledge Infrastructure\label{sec:ki}}
Knowledge Infrastructures (KI) are defined by Edwards~\cite{10.5555/1805940} as ``\textit{robust networks of people, artifacts, and institutions that generate, share, and maintain specific knowledge about the human and natural worlds}''.
Notable examples include the Intergovernmental Panel on Climate Change (IPCC) and the Wikipedia Community, with its individual contributors, formal and informal organizations, and (sub-)projects, digital libraries, tools, etc.
They are characterized by bringing together ``\textit{a diversity of actors, organizations and perspectives from, for instance, academia, industry, business and general public}''~\cite{karasti_knowledge_2016}, enduring beyond any individual project time~\cite{karasti_infrastructure_2010}.
As an emerging scientific field, it bridges the gap between different audiences and demographics, leading to a natural adjacency to another domain of open science: Science Communication.

\subsection{Science Communication\label{sec:scicom}}
Burns et al.~\cite{burns_science_2003}
define Science Communication (SciCom) as ``\textit{the use of appropriate skills, media, activities, and dialogue to produce one or more of the following personal responses to science (the AEIOU vowel analogy): Awareness, Enjoyment, Interest, Opinion-forming, and Understanding}''.
It aims to improve public understanding, acceptance, trust, and support while providing a venue to gather broad feedback, local knowledge, and civic needs regarding valuable research aims and applications~\cite{kappel_why_2019}.
Videos and podcasts, in particular, are widely used as knowledge distribution media.
MacKenzie~\cite{mackenzie_science_2019} has curated 952 science podcasts, \href{https://wissenschaftspodcasts.de/podcasts/}{wissenschaftspodcasts.de} 386; Kikkawa et al.~\cite{kikkawa_enhancing_2024,kikkawa_ya_2024} 230,000 videos, the world lecture project\footnote{\url{https://world-lecture-project.org/}} 59,634, the TIB-AV Portal~\cite{marin_arraiza_tibav_2015} 50,690.
These curators and creators are each actors in the SciCom KI.

\subsection{Science Communication Knowledge Infrastructure\label{sec:scicomki}}
The SciCom KI is hence the knowledge infrastructure dedicated to science communication, with its potentially fragmented individual actors, including people, artifacts, and institutions. 
Notable contributors include the Science Media Center (SMC) Global Network\footnote{\url{https://www.sciencemediacentre.org/international-smcs/}}, their national chapters, and individual members.
Sharing and maintaining knowledge are inherent to SciCom KI, which utilizes textual and non-textual artifacts to derive, preserve, and convey knowledge~\cite{kulczycki_transformation_2013}.
Particularly, videos are widely used and meticulously researched, with hundreds of different qualities and properties being analyzed and annotated~\cite{navarrete_closer_2023}.
This secondary data is usually stored in isolated repositories, disconnected from the original media, unless it is captured within FAIR infrastructure -- for instance, through licensing media in the TIB AV-Portal~\cite{marin_arraiza_tibav_2015} or by meeting the notability criteria\footnote{\url{https://www.wikidata.org/wiki/Wikidata:Notability}} of Wikidata~\cite{vrandecic_wikidata_2014}.
Yet, capturing this knowledge is especially relevant when assessing information that coexists within the same communication channels, but varies in quality, purpose, scope, evaluation standards, accessibility, and revenue generation~\cite{fahnrich_exploring_2023,hagenhoff_neue_2007}.

\begin{openboxwithtitle}{SciCom KI}
    ~\textbf{SciCom KI} is the \textbf{Knowledge Infrastructure} dedicated to \textbf{Science Communication};
    a network of \textbf{people} (e.g. content creators or curators), \textbf{artifacts} (e.g. platforms or data), and \textbf{institutions} (e.g. media outlets or research facilities).
\end{openboxwithtitle}

Evidently, the SciCom KI is very active, yet highly fragmented, lacking a common (knowledge) infrastructure, especially dedicated to audiovisual content.
Efforts exist, but are not unified in a collaborative environment such as the Wikiverse, which itself cannot meet these needs either.
Such a collaborative, platform-agnostic, linked open data infrastructure appears to be a gap in the state-of-the-art SciCom KI.

\section{Approach\label{sec:approach}}
To consolidate the fragmented SciCom KI efforts to curate NTM, we formalize a mission goal, elicit requirements via stakeholder survey, and design a system to fulfill these needs.

\begin{openboxwithtitle}{Mission Goal}
    ~Develop a platform to index information on scientific videos and podcasts that provides free access and participation for users to create and curate content alongside the principles of open source and open access.
\end{openboxwithtitle}

\subsection{Requirements Elicitation}
\begin{figure*}[t!]
    \centering
    \includegraphics[width=1\linewidth]{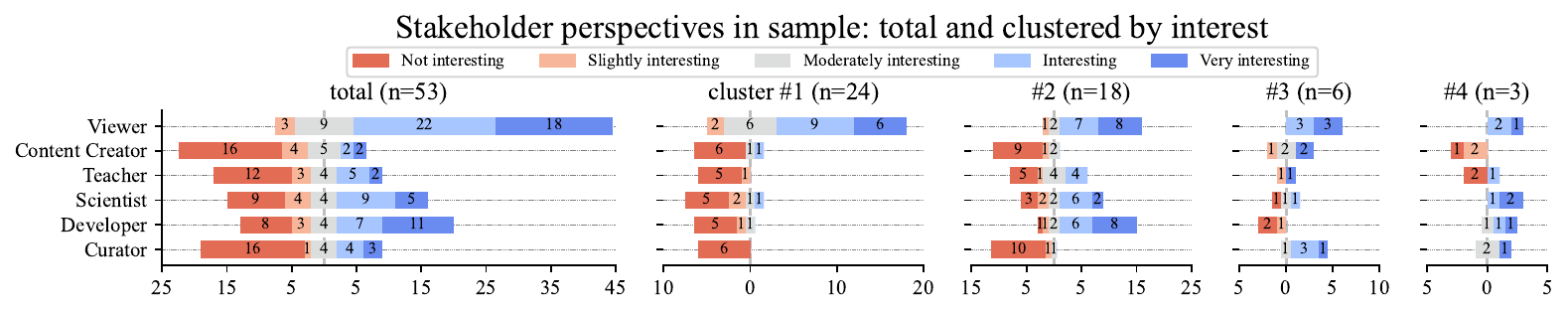}
    \caption{Total and clustered reported interest in our sample (n=53). The clusters indicate many participants are solely consumers, a second larger group can be classified as scientists and developers, a third as content creators and curators. Notably, almost every participant is interested as a viewer, often strongly.}
    \label{fig:perspective}
\end{figure*}
Given our mission goal, we conducted a domain analysis to identify a set of intermediate requirements and six key stakeholder roles within the SciCom KI: viewer, researcher, teacher, content creator, curator, and developer.
To accurately address and prioritize which criteria and features matter most to these stakeholders, we designed a survey.
The entire survey is available online\footnote{\url{https://github.com/borgnetzwerk/2025-scicom-ki-survey}}, including its results and evaluation.
It inquired about the individual perspectives regarding: 
which video or podcast libraries they use;
if they are interested in these media as
viewer,
content creator,
educator,
scientist,
developer or
curator; and
what troubles them most when searching for trustworthy videos and podcasts.
It then asked them to rank which tasks, criteria, and features the platform should focus on.
The tasks were
\textit{Find},
e.g., find the right podcast for your questions;
\textit{Curate},
e.g,. sort videos by topic, language or educational level;
\textit{Compare},
e.g,. show statements from different videos and podcasts in an overview; and
\textit{Debate},
e.g., support theories with arguments from videos and discuss open points.
The inquired features and criteria are listed in the following subsection alongside their results.

\subsection{Results}
\begin{figure}[tb]
    \includegraphics[width=1.0\linewidth]{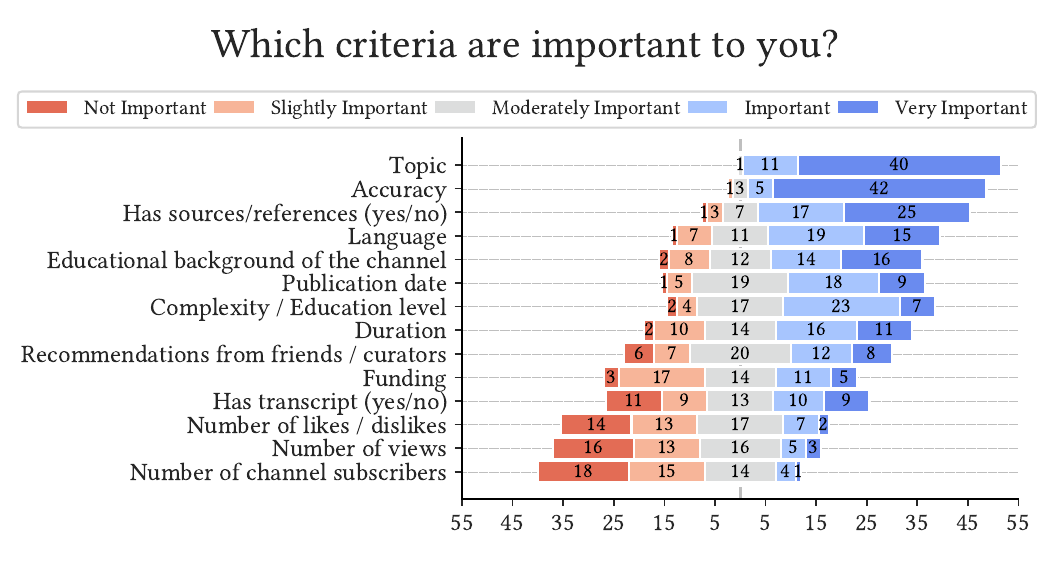}
    \includegraphics[width=1.0\linewidth]{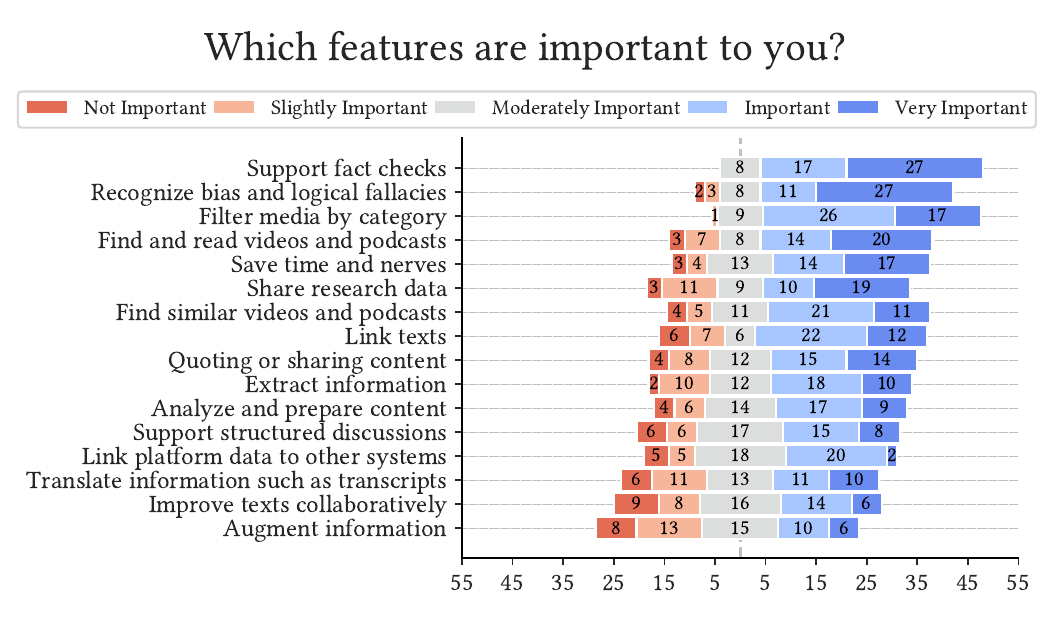}
    \caption{Results of 53 participants assessing the importance of Sci KI media criteria (above) and features (below), ranked by their importance averaged over all responses.}
    \label{fig:caf_ranking}
\end{figure}

The survey was distributed digitally, via flyer, and presented at the MediaWiki Users and Developers Conference~\cite{wittenborg_scicom_2024}.
Within one month, 53 participants had filled out the survey.
Their backgrounds are represented in \secref{fig:perspective}, the platforms used in \secref{tab:platforms}, and the ranking of criteria and features in \secref{fig:caf_ranking}.
\begin{table}[bth]
    \centering
    \caption{Number of participants that reported using these platforms. YouTube and Spotify dominate this space in our sample.}
    \begin{tabularx}{\linewidth}{X|c}
        \textbf{platform} & occurrences \\
        \midrule
        YouTube & 45 (85\%) \\
        \hline
        Spotify & 25 (47\%) \\
        \hline
        public broadcasting media library (e.g., ZDF or ARD) & 7  (13\%) \\
        \hline
        Amazon Prime Video, Apple Podcasts, Netflix & 5 each (9\%) \\
        \hline
        Audible & 4 (8\%) \\
        \hline
        Disney+, TikTok, Vimeo, Pocket Casts & 2 each (4\%) \\
        \hline
        \scriptsize
        Bosch Tube, Castbox, Castro Podcast app, Crunchyroll, Dailymotion, Deezer, Dropout, Facebook, Foss.video, Google, Instagram, LinkedIn, Mediathek, Nebula, Online Public Access Catalog (OPAC), Overcast, Podimo, Reddit, TIB AV-Portal, TIDAL, Tildes.net, Twitch, world-lecture-project.org & 1 each (2\%)
    \end{tabularx}
    \label{tab:platforms}
\end{table}

Despite our sampling being tailored towards users already interested in audio and video and not targeting an American audience, our findings of 85 \% using YouTube aligns with the Pew Research Center findings from 2024\footnote{\url{https://www.pewresearch.org/short-reads/2025/02/28/5-facts-about-americans-and-youtube/sr_25-02-28_youtube_1/}}.

Generally speaking, \textit{finding} media was the most important task (72\% \#1, next 9\% shared between discussing and comparing), followed by \textit{comparing} in second place (49\% \#2, next 17\% shared between discuss and curate) and \textit{curating} as third most important (40\% \#3, next 28\% for discussing).
Our survey included three open questions for missing criteria, features, or general suggestions.
19 participants provided additional input using these fields, which can be clustered as shown in \secref{tab:stakeholder_features}.

\begin{table*}[t!]
    \centering
    \caption{Additional Features and criteria suggested by the stakeholders, clustered by their common themes.}
    \begin{tabularx}{\textwidth}{l|X}
        \multicolumn{2}{l}{\textbf{Additional Annotation Propoerties}}\\
        \bottomrule
        Topicality, reliability, and neutrality & Assess to what degree content is up-to-date, reliable, and presented from a neutral standpoint.\\
        \hline
        Involvement in pseudoscience promotion & Disclaim when and how content or creators were advancing pseudoscientific claims.\\
        \hline
        (Undeclared) conflicts of interest & Identify potential undisclosed relationships of involved actors that may affect the content.\\
        \hline
        Advertisement / (Self-)promotion & Capture presence of advertisements and endorsements, including (self-)promotion.\\
        \hline
        Motivation and goal & Indicate the intent behind the content, such as education, entertainment, or economic.\\
        \hline
        Style (e.g., animation, commentary, satirical) & Describe the presentation style, including visually, narratively, as well as content-based.\\
        \hline
        Audio/video quality & Annotate technical quality, such as bitrate, resolution, or explicitly issues in general.\\
        \hline
        Vocal color & Note characteristics of the speakers' voices, such as low, warm, or raised voice.\\

        \toprule
        \multicolumn{2}{l}{\textbf{Inclusivity and Ethics}}\\
        \bottomrule
        Implement anti-discrimination measures & Flag discriminatory content, and provide resources for further information.\\
        \hline
        Advocate inclusive language & Highlight where inclusive language is used in content, but also in the system.\\
        \hline
        Include speaker details & Provide context and background information, including gender identity and disability status.\\
        \hline
        Focus on emotion and its influence on content & Describe emotional tone and its potential impact on message delivery.\\
        \hline
        Transparent and optional recommendation system & Provide optional recommendations with explanations of how they were identified.\\

        \toprule
        \multicolumn{2}{l}{\textbf{Collaboration and Trust}}\\
        \bottomrule
        Enable audience rating & Allow users to crowdsource various metrics, providing community-based assessments.\\
        \hline
        Implement a Web of Trust & Enable users to mark trusted peers and build trust networks, providing indirect trust indicators.\\
        \hline
        Support fact-checkers (e.g., Mimikama) & Facilitate and integrate the work of existing fact-checkers.\\
        \hline
        Support plugins (e.g., clickbait-remover) & Provide data for and ingest from new and existing plugins.\\
        \hline
        Cater to volunteers & Empower volunteers to utilize their motivation and meaningfully contribute.\\

        \toprule
        \multicolumn{2}{l}{\textbf{Accessibility and Usability}}\\
        \bottomrule
        Bookmarks & Enable users to create their own bookmarked media library within the system.\\
        \hline
        Watch history & Record previously viewed content for easier navigation and recommendations.\\
        \hline
        Keywords & Supports searching and filtering by relevant terms or topics.\\
        \hline
        Downloadable presentations/slides/blackboard images & Add references to supplementary materials used in videos.\\
        \hline
        Jump to the timestamp of specific words & Allow users to quickly locate and access segments of interest.\\
        \hline
        Part of a series & Indicate whether content is part of a larger collection or series.\\
        \hline
        Series details & Display information about the series, such as status, episode count, and last update.\\
    \end{tabularx}
    \label{tab:stakeholder_features}
\end{table*}

In summary, all participants had very individual requests, with several common threads.
Evidently, there is a strong desire to find accurate, topic-specific content in understandable language.
Likewise, participants seek trustworthy and transparent quality indicators, and improved accessibility and inclusivity features.
In that regard, it appears almost irrelevant if a medium in question previously had significant reach or publicity, so long as it is accurate and provides its sources.
Many suggestions also raise legal and ethical questions, such as annotating personal data or special categories thereof (e.g., gender identity), which warrant further investigation.
Our work focuses on the common threads and proceeds to implement an infrastructure meeting these needs.

\begin{openboxwithtitle}{Stakeholder Perspective}
~A platform to index information on scientific videos and podcasts should focus on making content from various platforms uniformly accessible.\\
The most important task to be supported is \textit{finding} media, followed by comparing and \textit{curating}, where \textit{discussing} is least important.\\
Several different filtering options are necessary, such as content-scoping (e.g., topic, language, date) and quality-indicators (e.g., accuracy, has sources, educational background). Popularity is apparently a very low priority.\\
A feature to support fact checks, alongside recognizing bias and logical fallacies, is unequivocally requested. No proposed feature was deemed unimportant overall, providing a wide variety of further development needs.
\end{openboxwithtitle}

\section{Implementation\label{sec:implementation}}
\begin{figure}[h!]
    \centering
    \includegraphics[width=1.0\linewidth]{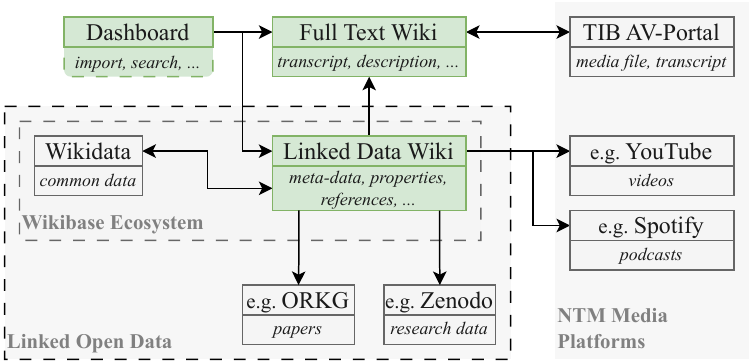}
    \caption{Overview of our proposed system (green) to connect knowledge from existing systems. The Dashboard serves as a non-technical, user-friendly entry point to the data curated in the Linked Data and Full Text Wikis.}
    \label{fig:overview}
\end{figure}

Wikidata~\cite{vrandecic_wikidata_2014} is conceptually capable of meeting most of the requested features, especially if combined with a capable long-text annotation platform such as Wikisource.
However, both of these existing platforms are already operating at capacity while addressing just notable, public domain content.
This was confirmed by several developers from inside and outside the Wikimedia team, including representatives from the Wikibase.cloud team, at the aforementioned MediaWiki Conference.
Hence, we extended the approach from MiMoText~\cite{schoch_smart_2022} by integrating our \nameref{sec:linked_data_wiki} into the Wikibase Ecosystem while expanding to a \nameref{sec:full_tex_wiki}, where legally permitted.
Our proposed system is highlighted in green in \secref{fig:overview}.
This open foundation was expanded with several microservices unified in a \nameref{sec:dashboard}, including a search page as well as content creation and aggregation automation.
All implemented services are illustrated in detail in the master thesis underlying this article~\cite{stehr_digitale_2025} and are open-source\footnote{Web-Page: \url{https://github.com/borgnetzwerk/dashboardduck}}\footnote{Media Search: \url{https://github.com/borgnetzwerk/searchsnail}}\footnote{Integration: \url{https://github.com/borgnetzwerk/integrationindri}}.
\begin{figure*}[b!]
    \centering
    \includegraphics[width=0.8\linewidth]{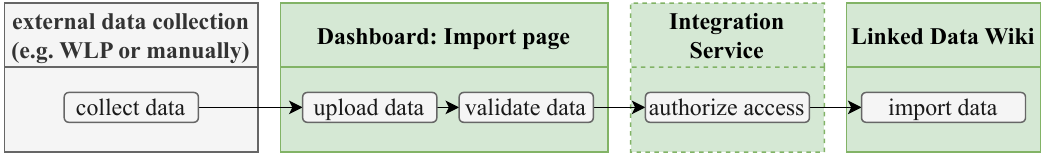}
    \caption{Data import facilitated by the Dashboard import page and the backend integration service. Content can be added and edited through the MediaWiki Frontend as well, but especially bulk ingestion is facilitated through the Dashboard. The MediaWiki API provides authorization and an edit history as well.}
    \label{fig:import}
\end{figure*}
\begin{figure*}[b!]
    \centering
    \includegraphics[width=0.8\linewidth]{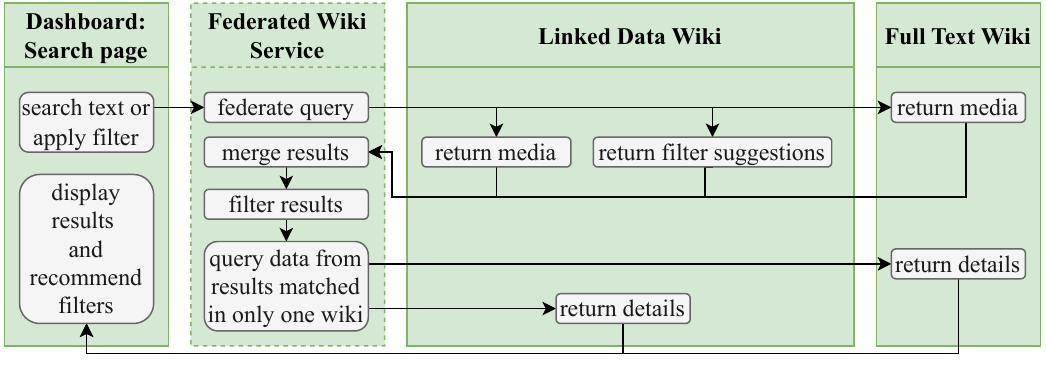}
    \caption{Process overview of the search page and its Federated Wiki Service. Query results are fetched from each wiki, filtered, and entries that are only matched in one wiki are complemented with the information from the respective other wiki. These results and suggested filters are then displayed.}
    \label{fig:search}
\end{figure*}

\subsubsection{Full Text Wiki\label{sec:full_tex_wiki}}
The Full Text Wiki fulfills two roles: first, storing long-form content such as video and podcast transcripts with tens of thousands of characters; second, providing a space not only to search, but also to annotate, improve, translate, semantify, and further process these textual representations.
Copyright restrictions limit the automated collection, processing, and particularly publication of transcripts.
Since the goal of our platform is to make information on audiovisual media more accessible, one of the main benefits is to have the textual transcript findable and available for further processing.
Yet, even if explicit author consent is given, transcripts may contain personal data, raising data protection concerns, which require a different type of consent far beyond the curator's or creator's capacity.
Before legally and ethically suitable solutions are found, our only available approach is to transcribe and ingest only content to which the copyright holder granted access and that has been reviewed for data protection compliance.
Hence, our work included a manually reviewed sample provided by Doktor Whatson\footnote{\url{https://www.youtube.com/@DoktorWhatson}}.

\begin{figure*}[bt!]
    \centering
    \includegraphics[clip,trim={0.5cm 8.6cm 10cm 11cm},width=0.49\linewidth]{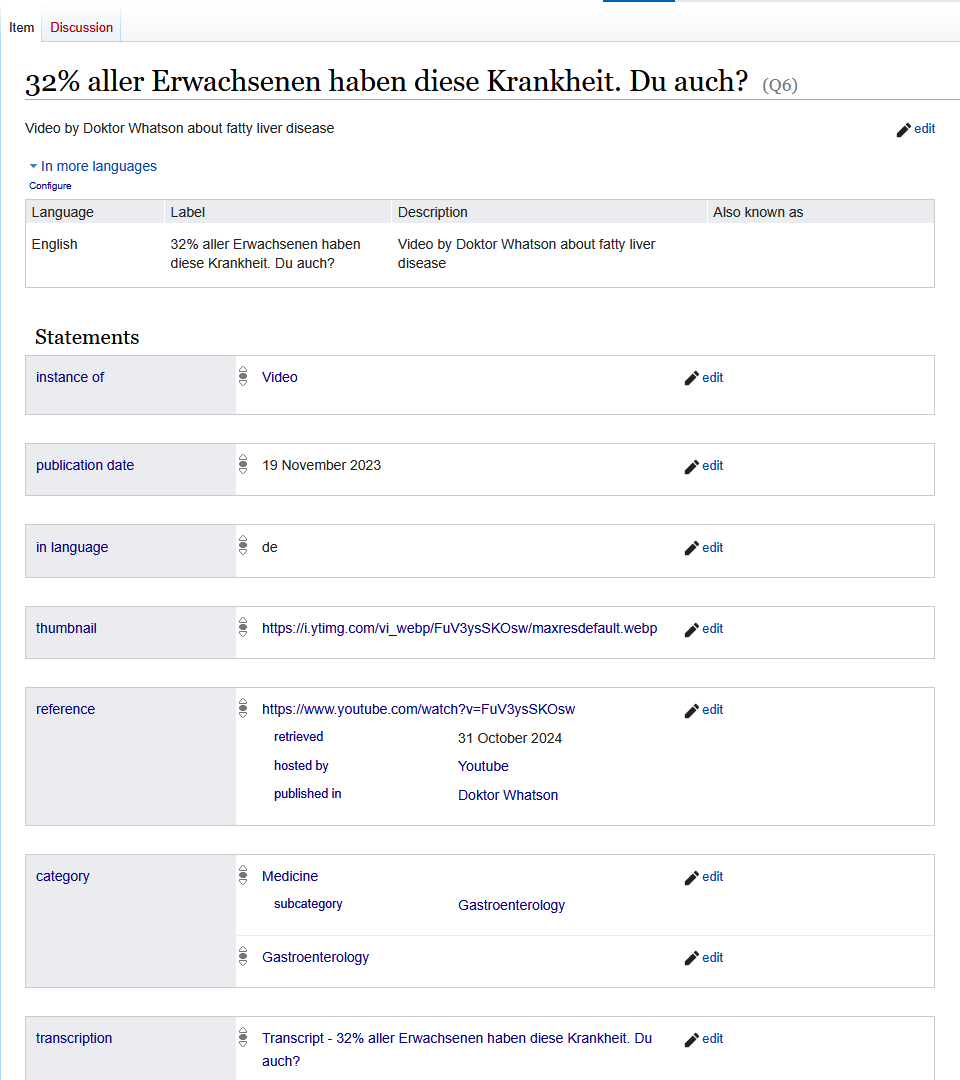}
    \includegraphics[clip,trim={1.5cm 13.5cm 0cm 1.75cm},width=0.49\linewidth]{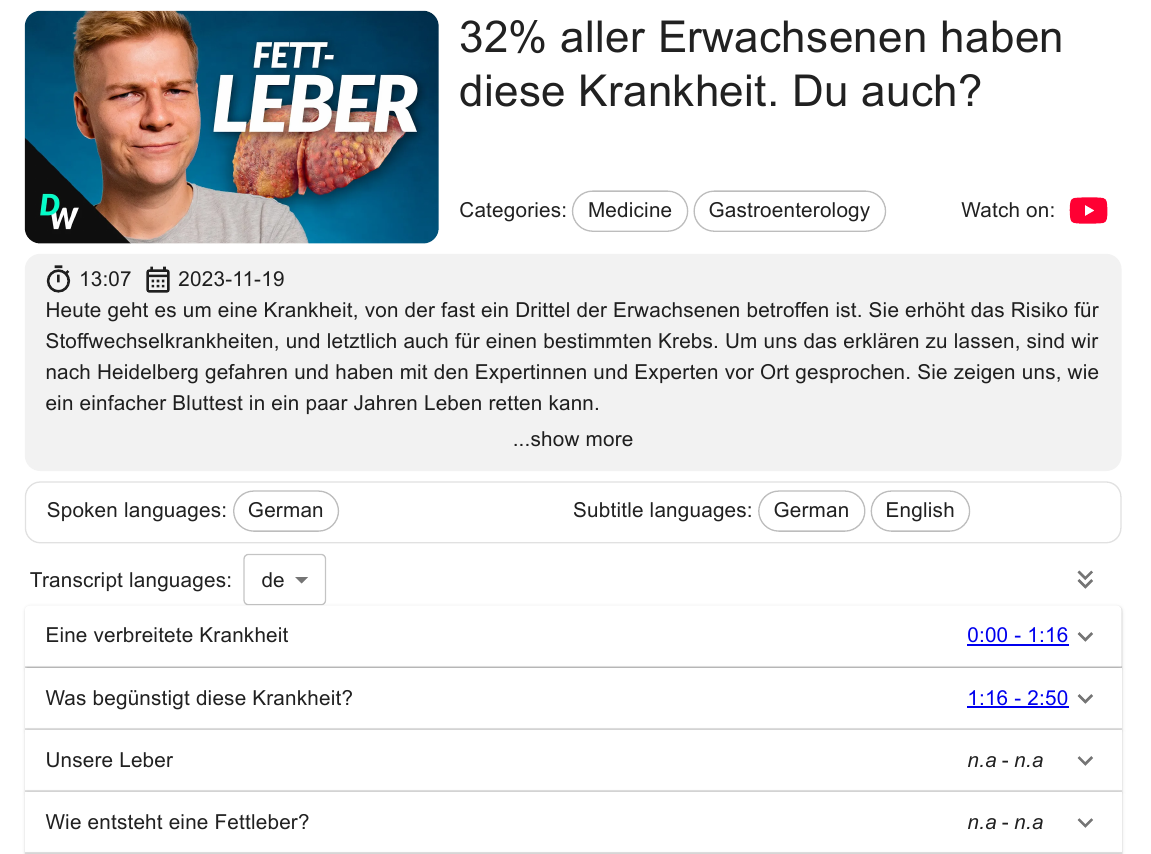}
    \caption{Knowledge graph representation of a media item on Wikibase (left), accessed by the Dashboard and displayed as a detail page (right). The Dashboard aggregates content from the references on the Wikibase, including the description and transcript from the Full Text Wiki.}
    \label{fig:detail_page}
\end{figure*}

\begin{figure*}[tb!]
    \centering
    \includegraphics[clip,trim={0cm 14.2cm 6.9cm 0cm},width=0.95\linewidth]{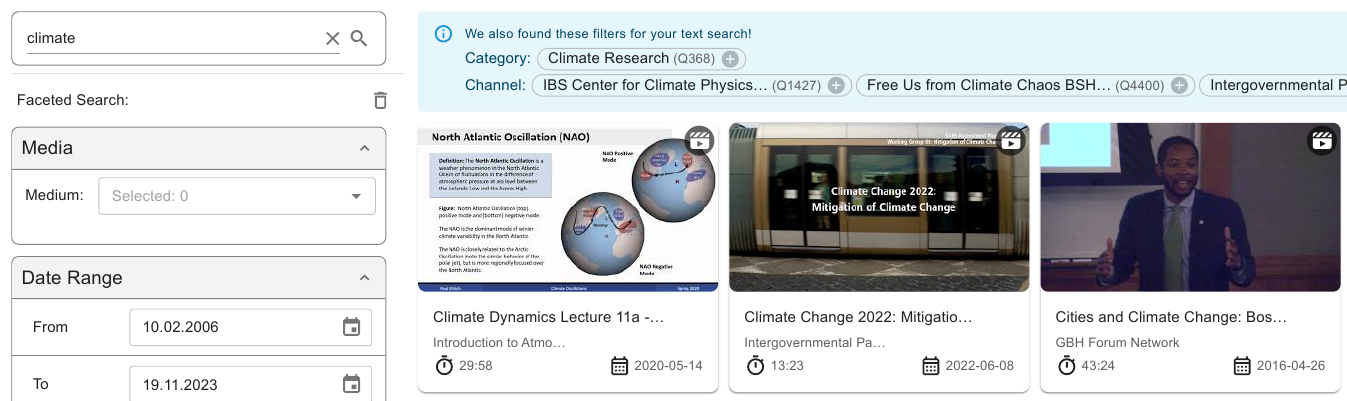}
    \caption{View when searching for climate-related media, including filter recommendations. The text is matched in both wikis, finding textual matches in the Full Text Wiki's artifacts as well as categorical matches in the Linked Data Wiki.}
    \label{fig:text_search_hint_cutout}
\end{figure*}

\subsubsection{Linked Data Wiki\label{sec:linked_data_wiki}}
The primary persistent layer of the infrastructure is a Wikibase installation.
Building on the overview of scientifically relevant media characteristics by Navarrete et al.~\cite{navarrete_closer_2023}, we identified over 200 different metadata qualities.
Examples include 
textual elements (e.g., transcripts, on-screen text) or
instructor behavior (e.g., tone, lexical diversity).
We complemented properties regarding 
FAIR context (e.g., license, accessibility),
involved actors (e.g., moderators, sponsors) or 
sources provided (e.g., on-screen, in description).
It serves as the central access point, as already illustrated in \secref{fig:overview}, responding to queries and linking to other data sources, such as the Full Text Wiki.

\subsubsection{Dashboard\label{sec:dashboard}}
When scaling to millions of heterogeneous artifacts, automation must be modular while ensuring content veracity.
For this purpose, we implemented an import service prototype that supports ingesting media metadata.
One such example is a dataset provided by the World Lecture Project\footnote{\url{https://world-lecture-project.org/}}, from which we created over 5,000 video representations with additional data enriched mainly from the YouTube API.
The process is briefly illustrated in \secref{fig:import}.
This service is designed primarily to support advanced users, who are engaging with the system at a deeper level beyond finding and filtering content via the search page.
The search page is designed as the core interface and entry point for most stakeholders, bringing together all systems.
As a view-only interface, it requires no login, reduces the complexity, and is hence designed to provide accessible usability for any user.
\Secref{fig:search} details the backend, allowing for properties and content stored in the Wikis to be used to filter for relevant media, as illustrated in \secref{fig:detail_page} and \ref{fig:text_search_hint_cutout}.

\section{Evaluation\label{sec:evaluation}}
The usability of the search page was examined via an online experiment.
Participants were acquired via a second wave of outreach similar to the previous survey, where respondents could already indicate their availability for future updates and interviews.
These online interviews were designed to evaluate satisfaction, objective and subjective efficiency and effectiveness, and the extent to which the implemented features satisfy the identified needs.

\subsection{Design}
Participants were given five tasks designed to test different functionalities and criteria.
These included searching for: 
\begin{enumerate}
    \item a video by a given title, 
    \item any video on ``history'' from the University of Göttingen, 
    \item any video on ``fatty liver'' published after 2022, 
    \item any video on ``computer science'' which is longer than 60 minutes and published in 2013 or 2014, and 
    \item any video in English.
\end{enumerate}
Task completion time was recorded to measure objective efficiency, while the number of correctly completed tasks determined effectiveness.
Subjective perceptions of effectiveness, efficiency, and provided support information were captured using an After-Scenario Questionnaire (ASQ) according to Lewis~\cite{lewis_psychometric_1991}, which uses a 7-point Likert scale from 1 (strongly agree) to 7 (strongly disagree).
User satisfaction was measured via the User Experience Questionnaire (UEQ)~\cite{schrepp_construction_2017}, covering attractiveness, perspicuity, efficiency, dependability, stimulation, and novelty.
Participants also rated how well the requested features and criteria  were implemented using a 5-point Likert scale from 1 (very good) to 5 (insufficient).
The study tested four hypotheses:
Users, on average, 
\begin{enumerate}
    \item complete tasks in under three minutes ($H_{A, \text{ efficiency}}$),
    \item have more than three tasks completed successfully ($H_{A, \text{ effectiveness}}$),
    \item report a $\text{UEQ} > 0.8$ ($H_{A, \text{ UX}}$), and
    \item  rate at least four of seven top criteria "good" or better ($H_{A, \text{ criteria}}$).
\end{enumerate}

\subsection{Results}
\begin{figure}[!bt]
    \centering
    \includegraphics[width=1.0\linewidth]{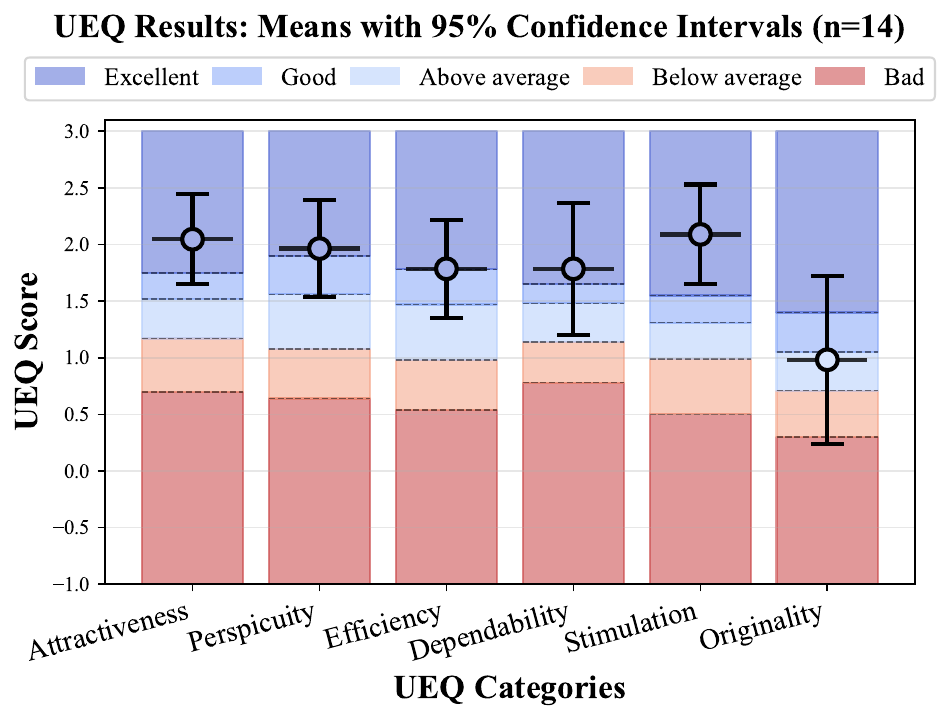}
    \caption{UEQ benchmark results across six UX scales. Notably, five of the means rank as "Excellent" and one "Above average": Originality. It has a high standard deviation ($\sigma=1.29$), with two participants ranking it negatively ($-0.75, -1.5$). Still, every aspect received overall positive feedback, confirming good or at least above average usability.}
    \label{fig:benchmark}
\end{figure}

\secref{tab:experiment_task_time} details the statistical tests for the hypotheses, confirming positive objective efficiency ($H_{A, \text{ efficiency}}$) and effectiveness ($H_{A, \text{ effectiveness}}$).
\begin{table}[b]
    \centering
    \caption{
    Statistical test results. The Shapiro-Wilk test assesses normality $N$ ($W, p$), determining the choice between t-test (for $N=yes$) and Wilcoxon Signed Rank ($N=no$). $t$- \& $p$-values are shown for t-tests; $Z$- and $p$-values for Wilcoxon tests. The null hypothesis ($H_0$) may be rejected based on the Bonferroni-Holm adjusted $p$-value ($p_a$).
    }
    \label{tab:experiment_task_time}
    \scriptsize
    \setlength\tabcolsep{2pt}
    \begin{tabular}{|c|ccc|cc|cc|c|c|}
    \hline
    \multirow{2}{*}{Efficiency} & \multicolumn{3}{l|}{Shapiro-Wilk Test} & \multicolumn{2}{c|}{t-Test} & \multicolumn{2}{l|}{\begin{tabular}[c]{@{}l@{}}Wilcoxon \\ Signed Rank\end{tabular}} & \multirow{2}{*}{$p_a$} & \multirow{2}{*}{\begin{tabular}[c]{@{}l@{}}reject \\$H_0$? \end{tabular}} \\ \cline{2-8}
    & \multicolumn{1}{c|}{W} & \multicolumn{1}{c|}{p} & N? & t & p & \multicolumn{1}{c|}{Z} & p & & \\ \hline
    Task 1 & \multicolumn{1}{c|}{0.861} & \multicolumn{1}{c|}{0.0316} & no & \multicolumn{1}{c|}{-} & - & \multicolumn{1}{c|}{-3.265} & 0.001 & 0.005 & yes \\ \hline
    Task 2 & \multicolumn{1}{c|}{0.873} & \multicolumn{1}{c|}{0.046} & no & \multicolumn{1}{c|}{-} & - & \multicolumn{1}{c|}{-2.711} & 0.007 & 0.0336 & yes \\ \hline
    Task 3 & \multicolumn{1}{c|}{0.87} & \multicolumn{1}{c|}{0.0423} & no & \multicolumn{1}{c|}{-} & - & \multicolumn{1}{c|}{-3.668} & $<$ 0.001 & 0.001 & yes \\ \hline
    Task 4 & \multicolumn{1}{c|}{0.68} & \multicolumn{1}{c|}{$<$ 0.001} & no & \multicolumn{1}{c|}{-} & - & \multicolumn{1}{c|}{-1.57} & 0.117 & 0.583 & no \\ \hline
    Task 5 & \multicolumn{1}{c|}{0.905} & \multicolumn{1}{c|}{0.1315} & yes & \multicolumn{1}{c|}{-26.179} & $<$ 0.001 & \multicolumn{1}{c|}{-} & - & $<$ 0.001 & yes \\ \hline
    \end{tabular}
    \label{tbl:effizienz}
\end{table}
\secref{fig:benchmark} illustrates the equally positive usability results ($\overline{{UEQ}} = 1.78$ ($H_{A, \text{ UX}})$).
Compared to the benchmark by Schrepp et al.~\cite{schrepp_construction_2017}, five categories scored excellent and one above-average.
The ASQ results were equally positive, with a mean effectiveness of 2.04 ($\sigma{} = 0.76$), efficiency of 1.67 ($\sigma{} = 0.68$), and provided support information of 2.075 ($\sigma{} = 0.64$).
Participants rated the implementation of three of the seven highest-priority criteria as “good” or better, namely topic, language, and release date, narrowly missing the $H_{A, \text{ criteria}}$ threshold of four.
Length and transcript availability were also rated as well implemented, yet they rank eighth and eleventh, respectively.
Other highly valued criteria, like correctness and complexity, are themselves complex to measure correctly, while some, like speaker background, face the personal data protection issues discussed earlier.
These challenges require dedicated scientific, legal, and ethical consideration before their metrics can be ingested and published.
While the current SciCom Wiki system does not yet fully satisfy all aspired features, it provides a consistently well-rated, usable, and conceptually extensible foundation for future work.

\begin{openboxwithtitle}{Current State}
~The current implementation of a Digital Library provides efficient and effective access to information on science communication videos and podcasts with above average to excellent usability ($\overline{{UEQ}} = 1.78$).\\
Of the seven identified most requested criteria, the current implementation achieved "good" or better satisfaction for three of them, leaving room for further improvement.
\end{openboxwithtitle}

\subsection{Threats to Validity}
We set out to develop a platform to index information on scientific videos and podcasts that provides free access and participation for users to create and curate content.
Our evaluation design can only capture the degree of success from a particular perspective, leaving limitations from construct to internal and external, as well as conclusion threats to validity.

\subsubsection{Construct Validity}
By design, our approach was sectioned into several stages:
The mission stage informed the stakeholder survey, which informed the implementation, based on which the evaluation was designed.
For example, our approach aimed to facilitate information curation and utilization, but the vast majority of interview stakeholders were more interested in the utilization aspect than in curation.
Due to several factors, our evaluation focused on this utilization aspect over curation:
Firstly, the underlying technology, most importantly MediaWiki, already facilitates the curation aspect, which needs no evaluation.
Secondly, while we have implemented semi-automated support for single- and bulk-data ingestion, using it requires several setup steps, including account- and API-key-creation, and was hence not included in the final evaluation addressed at the general user.
The structured data resulting from our curation automation was still the prerequisite of the experiment, which could not have worked without a set of curated data.
As such, our tool can support the creation and curation of content, but like Wikipedia, will primarily be used to utilize the knowledge\footnote{\url{https://en.wikipedia.org/wiki/Wikipedia:Statistics}}.
Even when focusing on utilization only, there is a chance that our selected task does not reflect the future user behavior, or that our measurements do not capture a relevant metric.
To address this, we used a selected set of tasks and evaluation metrics, broadening the scope of data captured, particularly ASQ for effectiveness and efficiency, UEQ for user experience.
By evaluating each aspect individually, we can make specific statements about different aspects of the application.

\subsubsection{Internal Validity}
During the interview, several gaps in our assumptions were revealed, skewing the metrics and validity of certain assumptions.
For example, we assumed a limit of three minutes for successful task completion would be a suitable proxy for measuring efficiency. Yet, some participants deliberately spent time exploring the application during the tasks, testing its limits and functionalities.
While filling out the user surveys, several users communicated that originality was neither expected nor necessarily positive, since examples such as YouTube and Wikipedia are so familiar that originality means deviation from what users are used to.
As such, each metric and measurement needs to be identified in its own context.
Similarly, social-desirability bias will likely have skewed both the initial survey as well as the interview, presumably the latter more than the former.
Due to these influences, the validity of statements such as "the popularity of a media is not important" and similar may be less stable and should be evaluated by other experiments less influenced by these biases.

\subsubsection{External Validity}
We have based our approach on a stakeholder survey of 53 participants, where our target audience is similar in generality to Wikipedia, reaching thousands of individuals per second.
As such, our sample is just the first starting point, requiring relentless further refinement to gather enough feedback to provide overall generalizability eventually.
The frequent use of free text fields similarly indicated that scoping the "video and podcast consumer" is a complex and nuanced task.
We could ensure that we captured these minute details in individual sessions, which, however, further reduced the participant numbers.
It is desirable that the upscaling of the proposed solution is tied to an upscaling of feedback inquiry, eventually reaching a status of generalizability to limit external threats to validity.

\subsubsection{Conclusion Validity}
The two major inferences that we conclude from our findings are that stakeholders desire a platform for more factual, transparent and accurate finding and comparing of videos and podcasts, and that our implementation provides an accessible working solution with potential for future improvement.
Particularly, the former is limited to the sampling of stakeholders, as previously highlighted, where the latter is limited to the tasks we explicitly tested.
To reach reliable conclusions on improving the SciCom KI through a digital library for science communication videos and podcasts, all aspects presented in this work and beyond must be adequately scaled.
Only by measuring impact alongside scaling the approach can reliable conclusions be drawn about the effect of a digital library solution.

\section{Discussion and Future Work\label{sec:discussion}}
We demonstrated a system to support the SciCom KI in some of the most requested identified features, such as finding, filtering, and structuring media.
Our digital library approach answers these stakeholder needs, which have confirmed efficiency, effectiveness, and usability. 
While our work has contributed to bracing against the flood of information, it has also shown that there is still an enormous amount of future work to be done.
This section discusses a selection of this future work, centered on our digital library approach, but also scoping the bigger picture in the Science Communication Knowledge Infrastructure.

\subsection{A digital library is just the first step}
We have established that information on videos and podcasts can be semantified, semi-automatically processed, stored and accessed on reliable, open-source infrastructure, enriched from various sources, and utilized to aid users in media discovery.
We have identified and explored various branches of improving this approach, including federated queries, dynamic filtering, usable yet transparent abstraction, and flexible yet reliable automation.
We have also identified optimization potential, such as strengthening the scalability of the federated query by setting up both wikis in a unified environment, allowing for direct access to their databases.
Yet, the insights gathered from our research, particularly from the survey and interview participants, indicate that even if we were to fully implement all feature requests we are currently aware of, this would still not meet the current needs of the SciCom KI.

\paragraph{Artifacts}
Science communication media is designed for human readability, not machine readability, yet machines facilitate its distribution and digestion.
The SciCom KI requires collaborative systems to meet its needs in an increasingly non-textual, factually ambiguous media landscape.
Our SciCom Wiki shares this goal of crowdsourcing audiovisual annotation with the world lecture project and advances it into the Wikibase Ecosystem.
This complements both Wikidata and TIB-AV-portal in scope, which cannot cover the required corpus breadth.
While this semantic advancement provides many new possibilities, the analysis of our in-person experiments suggests that general users seem to prefer systems that abstract away the underlying complexity and simplify the interaction.
While structured semantic data aids precision, it does not inherently improve accessibility for a broad audience.
Therefore, presenting this information in a user-friendly and easily digestible way remains a critical challenge.
Similarly, a digital library alone is not enough to stand out in an environment full of unreliable information.
Multimedia users will access the vast amount of their information not through our Dashboard, but through other interfaces.
The true potential of database systems stems not from their native interfaces, but from the heterogeneous applications built on top of them - from an overlay browser plugin to being interconnected in larger systems, such as Wikipedia being integrated into ChatGPT.
These derivative artifacts need to be consistently considered, encouraged, and supported towards a transparent, interconnected service environment.
Our Dashboard is just a demonstrator, one of several potential artifacts that utilize our proposed digital library.
The main advantage is the knowledge base, providing a collaboratively curated and interoperably accessed digital backbone for accessible interfaces to cover current media in everyday public discourse, providing tangible benefits to individual people.

\paragraph{People}
The SciCom KI requires active individuals in multiple roles: 
Viewers, researchers, teachers, content creators, curators, developers, and likely many more.
Viewers require several of the features we provided, but also structured venues to make their future feedback heard, raising concerns and requests to developers, for example.
Developers need a collaborative environment to address these, like the Wikimedia Phabricator\footnote{\url{https://phabricator.wikimedia.org/}}, to develop tools not only for viewers, but also for curators to more easily ingest, restructure, validate, complement, and curate data.
Content creators will likewise have their own perspective on what data is desired and lastly permitted to be processed in the first place.
Even with copyright permission, the legal limitations of data protection raise challenges for legal scholars to solve.
Particularly, these legal and political aspects elevate the success conditions to an institutional level.

\paragraph{Institutions}
While institutions provide a bigger environment for actors to collaborate within, they also provide a bigger target to attract opposition.
For example, even the curation of common domain content has already brought up legal struggles, like the recent data protection lawsuit against openJur\footnote{\url{https://openjur.de/u/2517464.html}}.
Any approach of this scale needs to take legally suitable precautions.
With the SciCom KI's scope exceeding international borders, these legal and ethical boundaries are even more nuanced and require their own dedicated research.
Just as we have highlighted that individual acceptance is fundamental to the success of our approach, institutional acceptance is equally required.
Only if all three components are considered - digital artifact prowess, individual people's needs, and institutional systemic boundaries - can we truly realize a digital library to support the science communication knowledge infrastructure for videos and podcasts.

\subsection{Future Work towards a robust SciCom KI}
Scholars could rightfully argue that the SciCom KI is currently too fragmented to be qualified being named as such, not passing Edwards' requirement of being a "robust network"~\cite{10.5555/1805940}.
Instead of arguing semantics, we focus on future work towards establishing a robust SciCom KI:

\begin{itemize}
    \item Further investigate lossless abstraction of complex information to provide context without contributing to the information overload.
    Our findings indicate that some complexity, such as the difference between a textual or categorical search or the presence of a wikibase-Q-identifier, may lead to unexpected confusion.
    \item Expand research and development towards meeting the SciCom KI stakeholder requirements, especially dedicated to individual roles.
    We saw that different actors bring different mental models~\cite{carroll_mental_1988}, which, like the roles themselves, are expected to shift over time.
    \item Establish project information platforms, such as Phabricator, a wiki, or Git, to maintain and advance projects.
    Building on the experience gathered from existing projects, especially the Wikiverse, these knowledge infrastructures within are of high importance to the "robustness" of the overarching systems.
    \item Connect people, especially contributors, with a low barrier to entry and in an accessible way, to capture low-level innovation potential.
    In the making of this work, dozens of individuals were contacted, many of whom are now connected in semi-structured collaborative environments. 
    These structures are often the difference between a fruitless idea and a successful collaboration.
    \item Ingest further data, but most importantly, onboard more people.
    It is always advised to get copyright holders on board before processing their data.
    Even if legally permissible, the ethical implications and public acceptance should always be considered.
    \item Build interoperable applications, and collaborate with (plugin-)developers.
    Many participants have requested and suggested various ways of interacting with the data, and the potential for integrating context into media consumption is vast once the underlying data is structured.
\end{itemize}

\begin{openboxwithtitle}{Discussion Summary}
~In the fragmented SciCom KI, our proposed system fills the gap of a repository for scalable crowd-sourcing towards FAIR audiovisual content and context representation.\\
To develop a \textbf{robust network} capable of generating, sharing, and maintaining knowledge on audiovisual science communication, all three KI aspects need to be further connected:\\
\textbf{Artifacts}, such as additional interfaces, collaborating platforms and ingested data,\\
\textbf{People}, from viewers providing feedback to contributing curators and involved creators, and various other roles,\\
\textbf{Institutions}, particularly media platforms and outlets, but also research facilities and low-level networks. 
\end{openboxwithtitle}

\section{Conclusion\label{sec:conclusion}}
This work addresses the increasing challenges of curating and utilizing FAIR (Findable, Accessible, Interoperable, Reusable) media within the emerging Science Communication Knowledge Infrastructure (SciCom KI).
Apparently, no existing digital library for information on scientific videos and podcasts scales to the rapid creation and dissemination of civically central information media.
We have presented and evaluated a federated, wiki-based digital library that enables scalable, collaborative representation of non-textual media, particularly videos and podcasts.

To inform the design of our digital library, we surveyed 53 stakeholders and interviewed another 11 for SciCom KI requirements, revealing a common, yet nuanced need from various perspectives to improve the findability and accessibility of audiovisual content and related context information.
Following Wikidata's Linked Open Data approach, we have designed and implemented a system to meet these needs.
A user-centered evaluation with 14 stakeholders in individual interviews confirmed its capacity to address several needs effectively.
This experiment was conducted online and allowed evaluators to explore the application while fulfilling five predefined tasks, measuring usability, effectiveness, and efficiency.
Participants reported good usability ($\overline{{UEQ}} = 1.78$) and equally positive ASQ results, with a mean effectiveness of 2.04 ($\sigma{} = 0.76$) and efficiency of 1.67 ($\sigma{} = 0.68$).

Beyond the findings related to approach filling an evident gap in the SciCom KI, our work identified a broader need for future work, including beyond improving technical infrastructure.
Other systems like Wikipedia and Wikidata have shown that providing a transparent, extensible, and community-driven platform lays the foundation for more reliable, accessible, and verifiable knowledge exchange.
Yet, these systems are evidently operating at capacity, at a much broader scope, and require complementary, domain-specific environments for more granular knowledge management.
The field of Science Communication requires similar infrastructure to empower researchers, educators, content creators, and especially citizens to address misinformation and foster trust in scientific knowledge collaboratively.
Further scaling, adoption, and expansion are necessary to develop the SciCom KI to meet all features requested by our identified stakeholders, and likely more to come once the scoping expands towards generalizability.
Our system demonstrated that a FAIR, collaborative, and user-friendly infrastructure for non-textual media is feasible, well-received by users, and lays the groundwork for more transparent, verifiable, and accessible science communication.
To realize the potential of the SciCom KI, we join the advice of our interviewed stakeholders and invite the research community, and also particularly volunteers, namely any person or institution interested in science communication media:
Collaborate to generate, share, and maintain knowledge on science communication videos and podcasts, in an openly accessible, transparent, and interoperable infrastructure as the system presented in this work, to address the challenges of the rapidly evolving information landscape and ultimately support informed civic discourse.

\section*{Acknowledgment}
\paragraph*{Use of AI tools declaration}
During the preparation of this work, the author(s) used \textbf{GPT-4.1 (GitHub Copilot)}, \textbf{DeepL}, \textbf{Grammarly (Browser Plugin)}, \textbf{LanguageTool (Browser Plugin)} in order to: \textbf{translate text}, \textbf{grammar and spelling check}, \textbf{paraphrase and reword}, \textbf{peer review simulation}, according to the CEUR GenAI Usage Taxonomy\footnote{\url{https://ceur-ws.org/GenAI/Taxonomy.html}}.
After using these tools/services, the authors reviewed and edited the content as needed and take full responsibility for the content of the publication.

\bibliographystyle{ieeetr}
\bibliography{main}

\end{document}